\documentclass[aps,prl,showpacs,twocolumn]{revtex4}%
\usepackage{amssymb}
\usepackage{amsfonts}
\usepackage{amsmath}
\usepackage{graphicx}
\usepackage{color}
\usepackage[dvips]{epsfig}
\usepackage[dvips]{graphicx}
\usepackage{float}%
\providecommand{\U}[1]{\protect\rule{.1in}{.1in}}
\begin{document}
\title{Radiative generation of the CPT-even gauge term of the SME from a
dimension-five nonminimal coupling term}
\author{R. Casana$^{a}$, M. M. Ferreira Jr$^{a}$, R. V. Maluf$^{b}$, F. E. P. dos
Santos$^{a}$, \thanks{e-mails: rodolfo.casana@gmail.com,
manojr07@ibest.com.br, robertovinhaes@fisica.ufc.br, fredegol@ibest.com.br.}}
\affiliation{$^{a}${\small {Universidade Federal do Maranh\~{a}o (UFMA), Departamento de
F\'{\i}sica, Campus Universit\'{a}rio do Bacanga, S\~{a}o Lu\'{\i}s - MA,
65085-580 - Brazil}}}
\affiliation{$^{b}${\small {Universidade Federal do Cear\'{a} (UFC), Departamento de
F\'{\i}sica, Campus do Pici, Fortaleza - CE, C.P. 6030, 60455-760 - Brazil}}}

\begin{abstract}
In this letter we show for the first time that the usual CPT-even gauge term
of the standard model extension (SME),\textbf{ }in its full structure, can be
radiatively generated, \textbf{ }in a gauge invariant level, in the context of
a modified QED endowed with a dimension-five nonminimal coupling term recently
proposed in the literature. As a consequence, the existing upper bounds on the
coefficients of the tensor $\left(  K_{F}\right)  $ can be used improve the
bounds on the magnitude of the nonminimal coupling, $\lambda\left(  K_{\sigma
F}\right)  ,$ by the factors $10^{5}$ or $10^{25}.$ The nonminimal coupling
also generates higher-order derivative contributions to the gauge field
effective action quadratic terms.

\end{abstract}

\maketitle

\section{Introduction}

During the last years it has been a great interest in theories endowed with
Lorentz symmetry violation. This interest was initially motivated by the
possibility of occurring this kind of violation in high energy theories
defined at the Planck energy scale \cite{Samuel}. The Standard Model Extension
(SME) \cite{Colladay} is the theoretical effective structure that includes
Lorentz-violating (LV) terms, generated as vacuum expectation values of
tensors quantities, in the different sectors of the usual Standard Model. A
large number of investigations in LV theories have been developed in recent
years, addressing distinct sectors of the SME: fermion systems \cite{fermion}%
,\ the CPT-odd gauge sector \cite{Adam,Cherenkov1,photons1}, the CPT-even
gauge sector \cite{KM1,Risse,Cherenkov2}. Interesting theoretical
generalizations involving higher dimensional LV operators have also been
devised \cite{Kostelec, Cambiaso, Myers}. These several studies have served
both to elucidate the effects engendered by Lorentz violation and to set up
stringent upper bounds on the LV coefficients \cite{KRussell}.

Another way to consider Lorentz violation effect in a usual physical theory is
by inserting new terms of interaction (LV nonminimal coupling terms) in the
Lagrangian. A pioneering study in this sense was undertaken in Ref.
\cite{NM1}, in which it was proposed a Lorentz-violating and CPT-odd
nonminimal coupling between fermions and the gauge field, $\displaystyle
D_{\mu}=\partial_{\mu}+ieA_{\mu}+i\frac{g}{2}\epsilon_{\mu\lambda\alpha\beta
}(k_{AF})^{\lambda}F^{\alpha\beta}$, in the context of the Dirac equation.
Here, $(k_{AF})^{\mu}$ is the Carroll-Field-Jackiw four-vector, and $g$ is the
constant that measures the nonminimal coupling magnitude, $\left[  g\right]
=mass^{-1}$. This nonrenormalizable theoretical proposal has been investigated
in several distinct tree-level scenarios
\cite{NM3,NMhall,NMmaluf,NMbakke,NMABC}. See also the references inside
\cite{NMABC}.

Another very interesting question examined in the literature is the radiative
generation of LV terms belonging to the SME framework. This topic was first
addressed in the end of 90's in Refs. \cite{Radio1}, in which it was argued
that the CPT-odd or Carroll-Field-Jackiw term, $\epsilon^{\mu\nu\rho\sigma
}\left(  k_{AF}\right)  _{\mu}A_{\nu}F_{\rho\sigma},$ belonging to the
electrodynamics of the SME is radiatively induced from the axial fermion LV
coupling, $b_{\mu}\bar{\psi}\gamma^{\mu}\gamma_{5}\psi$. This process leads to
the one-loop induced self-energy, $\Pi^{\mu\nu}=\kappa\epsilon^{\mu\nu
\alpha\beta}b_{\alpha}q_{\beta}$, whose coefficient\ $\kappa$ depends
explicitly on the regularization prescription used to control the UV
divergencies. Such ambiguity has unleashed a great controversy about the
possibility the CFJ term to be radiatively generated or not, since a gauge
invariant prescription,\ in principle, would provide $\kappa=0$ \cite{Radio2}.
Other developments in radiative generation,\textsl{ }including the induction
of the Chern-Simons-like action in a LV massless QED, finite temperature
effects, and photon triple splitting, were addressed in Ref. \cite{Radio3}.

In Ref. \cite{Gomes} it was demonstrated that aether-like \cite{Aether},
CPT-even, and Lorentz-violating terms can be properly generated when the
suitable couplings to the spinor fields are considered. In the first work of
Ref. \cite{Tiago2} the first higher derivative (dimension-five) CPT-odd
operator of the extended QED proposed in Ref. \cite{Kostelec} was radiatively
induced by the fermion sector term involving the coefficient\ $g^{\alpha\mu
\nu}.$ The radiative generation of other higher-dimensional gauge terms,
including the Myers-Pospelov one, was achieved in the second work of Ref.
\cite{Tiago2}, starting from a modified QED based on the presence of the
CPT-odd nonminimal\ coupling of Ref. \cite{NM1}. An interesting study about a
QED model also modified by the nonminimal coupling $\epsilon_{\mu\lambda
\alpha\beta}\gamma^{\mu}b^{\lambda}F^{\alpha\beta}$ was performed\ in Ref.
\cite{BaetaR1}, where it was shown that the one-loop quantum corrections to
the photon self-energy could provide two LV terms in the photon sector: the
usual CPT-odd Carroll-Field-Jackiw term and the CPT-even aether one,
$b_{\alpha}b_{\mu}F^{\alpha\beta}F^{\mu\nu},$ first generated in Ref.
\cite{Gomes}. This aether term may recover some components of CPT-even usual
term, $\left(  K_{F}\right)  _{\mu\nu\alpha\beta},$ but not its entire
structure. \ Note that the term $b_{\alpha}b_{\mu}$\ is just a piece of the
representation of the $\left(  K_{F}\right)  _{\mu\nu\alpha\beta}$\ in terms
of the vector $b_{\mu},$\ stated in Eq. (17) of Ref. \cite{BaetaR1}, which
maps only the nonbirefringent sector of the tensor $\left(  K_{F}\right)  .$
In Ref. \cite{BaetaR2}, a similar investigation was performed considering a
chiral version of this nonminimal coupling, that is, $\epsilon_{\mu
\lambda\alpha\beta}\gamma_{5}\gamma^{\mu}b^{\lambda}F^{\alpha\beta},$ with
analogous results. New LV radiatively generated terms were recently achieved
in Ref. \cite{BaetaR3}. Some clues about the radiative generation of the
nonbirefringent part of the CPT-even tensor, $(K_{F})_{\text{ \ \ }\mu\rho\nu
}^{\rho}=(c_{\mu\nu}+c_{\nu\mu})/2,$ from the fermion sector term, $c_{\mu\nu
},$ can be found in Ref. \cite{Kostelecky2002}, in which the 1-loop
renormalization of the SME Quantum Electrodynamics was demonstrated. Note,
however, that this contribution only generates a part of the full $\left(
K_{F}\right)  $ tensor.

In two very recent works \cite{Frede1,Frede2}, it has been proposed a new
CPT-even, dimension-five, nonminimal coupling linking the fermionic and gauge
fields, in the Dirac equation, $(i\gamma^{\mu}D_{\mu}-m)\Psi=0,$ where
$D_{\mu}$\ is a nonminimal covariant derivative,
\begin{equation}
D_{\mu}=\partial_{\mu}+ieA_{\mu}+\frac{\lambda}{2}\left(  K_{F}\right)
_{\mu\nu\alpha\beta}\gamma^{\nu}F^{\alpha\beta}, \label{Cov1}%
\end{equation}
in which $\left(  K_{F}\right)  _{\mu\nu\alpha\beta}$ is the CPT-even tensor
of Abelian gauge sector of the SME, and the nonminimal coupling constant
$\lambda$ has dimension mass$^{-1}.$ The corresponding fermionic Lagrangian
density is
\begin{equation}
\mathrm{{\mathcal{L}}}_{\Psi}=\bar{\Psi}\left[  i\partial
\!\!\!\slash-eA\!\!\!\slash-m+\frac{\lambda}{2}\left(  K_{F}\right)  _{\mu
\nu\alpha\beta}\sigma^{\mu\nu}F^{\alpha\beta}\right]  \Psi, \label{L1a}%
\end{equation}
with $\Psi$ representing a Dirac usual spinor, and $\sigma^{\mu\nu}=\frac
{i}{2}[\gamma^{\mu},\gamma^{\nu}].$

In Ref. \cite{Frede1}, one has studied the effects implied by this new term on
the cross section of the electron-muon scattering. The analysis of the
ultrarelativistic limit and the available experimental data has allowed to
attain the upper bound, $\left\vert \lambda\left(  K_{F}\right)  \right\vert
\leq10^{-12}\left(  \mbox{eV}\right)  ^{-1}$. On the other hand, the role
played by this nonminimal coupling on the nonrelativistic regime of the Dirac
equation was analyzed in Ref. \cite{Frede2}, focusing on new corrections
induced on the hydrogen spectrum and on the gyromagnetic constant. Such
analysis has implied an upper bound as restrictive as $\left\vert
\lambda\left(  K_{F}\right)  \right\vert \leq10^{-16}\left(  \mbox{eV}\right)
^{-1}$.

In this letter, we show for the first time the radiative generation of the
full CPT-even term of the SME electrodynamics, $\left(  K_{F}\right)  _{\mu
\nu\alpha\beta}F^{\mu\nu}F^{\alpha\beta}$, \ embracing the entire structure of
the tensor $\left(  K_{F}\right)  _{\mu\nu\alpha\beta}$. This is performed by
means of a gauge invariant way by starting from the nonminimal CPT-even
coupling (\ref{Cov1}) introduced in the Dirac equation. We finalize presenting
the second order contributions in the tensor $\left(  K_{F}\right)  $\ to the
photon self-energy. In the conclusions, we discuss how these results may
improve some previous upper bounds on the magnitude of the CPT-even nonminimal
coupling, yielding $\left\vert \lambda_{e}\left(  K_{F}\right)  \right\vert
<10^{-21}\left(  \text{eV}\right)  ^{-1}$\ and $\left\vert \lambda_{e}\left(
K_{F}\right)  \right\vert <10^{-41}\left(  \text{eV}\right)  ^{-1}$ for the
nonbirefringent and birefringent coefficients, respectively.

\section{Effective action}

The QED model under consideration is%
\begin{equation}
\mathcal{L}=-\frac{1}{4}F^{\mu\nu}F_{\mu\nu}+\mathrm{{\mathcal{L}}}_{\Psi},
\end{equation}
where $L_{\Psi}$ is given in Eq.(\ref{L1a}), and a convenient gauge fixing
term must be introduced to properly define the quantization procedure. From
now on we will change the nonminimal notation as
\begin{equation}
\lambda\left(  K_{F}\right)  _{\mu\nu\alpha\beta}\rightarrow\lambda\left(
K_{\sigma F}\right)  _{\mu\nu\alpha\beta},
\end{equation}
in order to loyally follow the idea that this term couples the gauge tensor
and the Dirac bilinear $\bar{\Psi}\sigma^{\mu\nu}\Psi$. Firstly, we are
interested in the contributions of the fermionic fields, undergone to the
nonminimal coupling interaction, to the effective action of the
electromagnetic field. The full contribution of the fermion fields to the
gauge field effective action is attained by integrating on the fermionic
field, yielding
\begin{equation}
e^{iW\left[  A\right]  }=\frac{\det\left(  i\partial\!\!\!\slash-B-m\right)
}{\det\left(  i\partial\!\!\!\slash-m\right)  }, \label{Effec2}%
\end{equation}
with the matrix $B$ operator defined as
\begin{equation}
B=eA\!\!\!\slash-\lambda\left(  K_{\sigma F}\right)  _{\mu\nu\alpha\beta
}\sigma^{\mu\nu}\partial^{\alpha}A^{\beta}. \label{B1}%
\end{equation}
We are interested in the quadratic term in the gauge field\ coming from\ Eq.
(\ref{Effec2}), which is equivalent to the second order contribution in $e$
and $\lambda$,\textbf{\ }written in the momentum space as%
\begin{equation}
W^{\left(  2\right)  }\left[  A\right]  =-\frac{1}{2}\int\frac{d^{4}q}{\left(
2\pi\right)  ^{4}}\tilde{A}_{\mu}\left(  -q\right)  \Pi^{\mu\nu}\left(
q\right)  \tilde{A}_{\nu}\left(  q\right)  , \label{Effec7}%
\end{equation}
\ where $\Pi^{\mu\nu}\left(  q\right)  $\ is the one-loop photon self-energy%
\begin{equation}
\Pi^{\mu\nu}\left(  q\right)  =-i\int\frac{d^{4}p}{\left(  2\pi\right)  ^{4}%
}\text{tr}\left[  \tilde{S}\left(  p\right)  V^{\mu}\tilde{S}\left(
p+q\right)  V^{\nu}\right]  . \label{SelfE0}%
\end{equation}
The symbol \textquotedblleft tr\textquotedblright\ denotes the trace operation
in the Dirac's indices, $\tilde{S}\left(  p\right)  $\ is the fermionic
Feynman propagator,
\begin{equation}
\tilde{S}(p)=i\left(  p\!\!\!\slash-m\right)  ^{-1}, \label{ferr}%
\end{equation}
and the quantities $\tilde{B}\left(  q\right)  $\ and $V_{\beta}\left(
q\right)  $\ are given by
\begin{align}
\tilde{B}\left(  q\right)   &  =V_{\beta}\left(  q\right)  \tilde{A}^{\beta
}\left(  q\right)  ,\label{B2a}\\[0.3cm]
V_{\beta}\left(  q\right)   &  =e\gamma_{\beta}+i\lambda\left(  K_{\sigma
F}\right)  _{\mu\nu\alpha\beta}\sigma^{\mu\nu}q^{\alpha}. \label{Vertex1}%
\end{align}

We can justify the introduction in Eq. (\ref{Effec7}) of the 1-loop photon
self-energy as part of the two-point component of the gauge field effective
action because it will allow to show that the radiative corrections preserve
the transversality condition, guaranteeing the gauge invariance at this level.

\subsection{One-loop vacuum polarization}

In order to evaluate the one-loop corrections to the photon self-energy,
expression (\ref{SelfE0}) is rewritten as the sum
\begin{equation}
\Pi^{\mu\nu}\left(  q\right)  =%
{\displaystyle\sum\limits_{(a,b)}}
\Pi_{(a,b)}^{\mu\nu}\left(  q\right)  ,
\end{equation}
where $\Pi_{(a,b)}^{\mu\nu}\left(  q\right)  $\ is defined by
\begin{equation}
\Pi_{(a,b)}^{\mu\nu}\left(  q\right)  =i\int\frac{d^{4}p}{\left(  2\pi\right)
^{4}}\frac{N_{\left(  a,b\right)  }^{\mu\nu}}{\left(  p^{2}-m^{2}\right)
\left(  \left(  p+q\right)  ^{2}-m^{2}\right)  }. \label{SelfE1}%
\end{equation}
It is obtained by replacing the fermion Feynman propagators (\ref{ferr}) in
Eq. (\ref{SelfE0}), so that\ $N_{\left(  a,b\right)  }^{\mu\nu}$\ is
\begin{equation}
N_{\left(  a,b\right)  }^{\mu\nu}=\text{tr}\left[  (p\!\!\!\slash+m)V_{\left(
a\right)  }^{\mu}\left(  p\!\!\!\slash+q\!\!\!\slash+m\right)  V_{\left(
b\right)  }^{\nu}\right]  ,
\end{equation}
with $a,b=0,1$\ representing the usual and modified vertices,
\begin{equation}
V_{\left(  0\right)  }^{\mu}=e\gamma^{\mu},\text{ \ \ }V_{\left(  1\right)
}^{\mu}=i\lambda\left(  K_{\sigma F}\right)  ^{\alpha\beta\chi\mu}%
\sigma_{\alpha\beta}q_{\chi}, \label{vertex2}%
\end{equation}
respectively. The integral in (\ref{SelfE1}) is a quadratically divergent by
power counting requiring some regularization technique, in our case, we will
use an explicitly gauge invariant prescription: the dimensional
regularization. The dimensional regularization works in $D=4-2\epsilon
$\ dimensions, regarding $\epsilon\rightarrow0^{+}$.

Hence, in our notation, $\Pi_{\left(  0,0\right)  }^{\mu\nu}$\ represents the
usual one-loop self-energy corresponding to the vertex $V_{\left(  0\right)
}^{\mu}$\ while the new contributions involving the new vertex\ $V_{\left(
1\right)  }^{\mu}$\ namely $\Pi_{\left(  0,1\right)  }^{\mu\nu},\Pi_{\left(
1,0\right)  }^{\mu\nu},\Pi_{\left(  1,1\right)  }^{\mu\nu}$.

The first contribution to be considered is the usual one-loop photon
self-energy contribution, $\Pi_{\left(  0,0\right)  }^{\mu\nu}$, read as%
\begin{equation}
\Pi_{\left(  0,0\right)  }^{\mu\nu}\left(  q\right)  =ie^{2}\int\frac{d^{4}%
p}{\left(  2\pi\right)  ^{4}}\frac{N_{\left(  0,0\right)  }^{\mu\nu}}{\left(
p^{2}-m^{2}\right)  \left(  \left(  p+q\right)  ^{2}-m^{2}\right)  },
\label{Pi00}%
\end{equation}
with
\begin{equation}
N_{\left(  0,0\right)  }^{\mu\nu}=\text{tr}\left[  (p\!\!\!\slash+m)\gamma
^{\mu}\left(  p\!\!\!\slash+q\!\!\!\slash+m\right)  \gamma^{\nu}\right]  .
\end{equation}
By following the dimensional regularization technique, we perform\ the trace
operations and compute the momentum integral with the Feynman parametrization.
Next, we retain\ only the contribution of the divergent terms,\
\begin{equation}
\Pi_{(0,0)}^{\mu\nu}(q)=-\frac{1}{12\pi^{2}\epsilon}\left(  g^{\mu\nu}%
q^{2}-q^{\mu}q^{\nu}\right)  , \label{pol1}%
\end{equation}
for the quadratic term of the gauge field effective action, given
as\textbf{\ }
\begin{equation}
W_{\left(  0,0\right)  }^{(2)}\left[  A\right]  =-\frac{e^{2}}{48\pi
^{2}\epsilon}\int d^{4}x\ F^{\mu\nu}F_{\mu\nu}. \label{EffA1a}%
\end{equation}
As expected, the usual vertex induces a counterterm proportional to the
Maxwell term, $F^{\mu\nu}F_{\mu\nu}$, which is already present in the QED action.

We now go on evaluating the terms $\Pi_{\left(  0,1\right)  }^{\mu\nu}%
,\Pi_{\left(  1,0\right)  }^{\mu\nu}$. A\ preliminary analysis allows to
notice that $\Pi_{\left(  0,1\right)  }^{\mu\nu}=\Pi_{\left(  1,0\right)
}^{\mu\nu}$, so we go compute the first one, in which the replacement of the
vertices (\ref{vertex2}) yields\textbf{ }%
\begin{equation}
\Pi_{(0,1)}^{\mu\nu}\left(  q\right)  =-e\lambda q_{\sigma}\left(  K_{\sigma
F}\right)  _{\alpha\beta}{}^{\sigma\nu}\Pi^{\mu\alpha\beta},
\end{equation}
\textbf{\ }where%
\begin{equation}
\Pi^{\mu\alpha\beta}=\int\frac{d^{4}p}{(2\pi)^{4}}\frac{N^{\mu\alpha\beta}%
}{(p^{2}-m^{2})\left[  (p+q)^{2}-m^{2}\right]  }, \label{Int01}%
\end{equation}
and
\begin{equation}
\ N^{\mu\alpha\beta}=\text{tr}\left[  \left(  p\!\!\!\slash+m\right)
\gamma^{\mu}\left(  p\!\!\!\slash+q\!\!\!\slash+m\right)  \sigma^{\alpha\beta
}\right]  .
\end{equation}
We calculate $\Pi^{\mu\alpha\beta}$\ by following the same procedure used to
compute the quantity (\ref{Pi00}), thus the divergent term is\textbf{\ }%
\begin{equation}
\Pi^{\mu\alpha\beta}=-\frac{m}{4\pi^{2}\epsilon}\left(  q^{\alpha}g^{\beta\mu
}-q^{\beta}g^{\alpha\mu}\right)  ,
\end{equation}
which yields the following contribution to the photon self-energy:%
\begin{equation}
\Pi_{\left(  0,1\right)  }^{\mu\nu}\left(  q\right)  =\frac{me\lambda}%
{4\pi^{2}\epsilon}\left(  K_{\sigma F}\right)  _{\alpha\beta}{}^{\sigma\nu
}q_{\sigma}\left(  q^{\alpha}g^{\beta\mu}-q^{\beta}g^{\alpha\mu}\right)  .
\label{pol2}%
\end{equation}
Inserting it in the effective action, one attains
\begin{equation}
W_{(0,1)}^{\left(  2\right)  }\left[  A\right]  +W_{(1,0)}^{\left(  2\right)
}\left[  A\right]  =\frac{me\lambda}{8\pi^{2}\epsilon}\int d^{4}x~\left(
K_{\sigma F}\right)  _{\mu\nu\alpha\beta}^{\text{ \ \ \ \ }}F^{\mu\nu
}F^{\alpha\beta}. \label{EffA2}%
\end{equation}
This result reveals that the CPT-even abelian gauge term of the SME, $\left(
K_{F}\right)  _{\mu\nu\alpha\beta}F^{\mu\nu}F^{\alpha\beta}$, is
radiatively\ induced by the new vertex, once the tensor $\left(  K_{\sigma
F}\right)  $ has exactly the same symmetries as $\left(  K_{F}\right)  .$ This
is the first time this full CPT-even term is generated by a gauge-invariant mechanism.

We finalize evaluating the term $\Pi_{\left(  1,1\right)  }^{\mu\nu}$, which
after vertex substitution\ can be rewritten as\textbf{ }%
\begin{equation}
\Pi_{(1,1)}^{\mu\nu}\left(  q\right)  =-i\lambda^{2}\left(  K_{\sigma
F}\right)  _{\eta\theta}{}^{\xi\mu}\left(  K_{\sigma F}\right)  _{\lambda\rho
}{}^{\chi\nu}q_{\xi}q_{\chi}\Pi^{\eta\theta\lambda\rho}\left(  q\right)  ,
\end{equation}
where we have defined\textbf{ }
\begin{equation}
\Pi^{\eta\theta\lambda\rho}\left(  q\right)  =\int\frac{d^{4}p}{\left(
2\pi\right)  ^{4}}\frac{N^{\eta\theta\lambda\rho}}{\left(  p^{2}-m^{2}\right)
\left[  (p+q)^{2}-m^{2}\right]  }, \label{n33}%
\end{equation}
and%
\begin{equation}
N^{\eta\theta\lambda\rho}=\text{tr}\left[  (p\!\!\!\slash+m)\sigma^{\eta
\theta}\left(  p\!\!\!\slash+q\!\!\!\slash+m\right)  \sigma^{\lambda\rho
}\right]  .
\end{equation}

Direct computation of the integral (\ref{n33}) allows to get the following
divergent terms\textbf{\ }%
\begin{align}
\Pi^{\eta\theta\lambda\rho}\left(  q\right)   &  =-\frac{i}{4\pi^{2}\epsilon
}\left(  m^{2}-\frac{q^{2}}{6}\right)  \left[  g^{\eta\rho}g^{\theta\lambda
}-g^{\eta\lambda}g^{\theta\rho}\right] \nonumber\\[-0.3cm]
& \\
&  \hspace{-1cm}-\frac{i}{12\pi^{2}\epsilon}\left[  q^{\eta}q^{\rho}%
g^{\theta\lambda}-q^{\eta}q^{\lambda}g^{\theta\rho}+q^{\theta}q^{\lambda
}g^{\eta\rho}-q^{\theta}q^{\rho}g^{\eta\lambda}\right]  ,\nonumber
\end{align}
providing the second-order LV contributions to the photon self-energy,%
\begin{align}
\Pi_{(1,1)}^{\mu\nu}\left(  q\right)   &  =-\frac{m^{2}\lambda^{2}}{2\pi
^{2}\epsilon}\left(  K_{\sigma F}\right)  ^{\mu\xi\lambda\rho}\left(
K_{\sigma F}\right)  _{\lambda\rho}{}^{\chi\nu}q_{\xi}q_{\chi}\nonumber\\
&  +\frac{\lambda^{2}}{12\pi^{2}\epsilon}\left(  K_{\sigma F}\right)  ^{\mu
\xi\lambda\rho}\left(  K_{\sigma F}\right)  _{\lambda\rho}{}^{\chi\nu}%
q^{2}q_{\xi}q_{\chi}\label{pol3}\\
&  +\frac{\lambda^{2}}{3\pi^{2}\epsilon}\left(  K_{\sigma F}\right)  ^{\mu
\xi\eta\lambda}\left(  K_{\sigma F}\right)  _{\lambda}{}^{\rho\chi\nu}q_{\xi
}q_{\eta}q_{\rho}q_{\chi},\nonumber
\end{align}
and the following counterterms to the effective gauge field action,
\begin{align}
W_{\left(  1,1\right)  }^{\left(  2\right)  }  &  =-\frac{\lambda^{2}m^{2}%
}{16\pi^{2}\epsilon}\int d^{4}x~\left(  K_{\sigma F}\right)  ^{\mu\xi
\rho\theta}\left(  K_{\sigma F}\right)  _{\rho\theta}{}^{\chi\nu}F_{\mu\xi
}F_{\chi\nu}\nonumber\\
&  \hspace{-1cm}+\frac{\lambda^{2}}{96\pi^{2}\epsilon}\int d^{4}x~\left(
K_{\sigma F}\right)  ^{\mu\xi\rho\theta}\left(  K_{\sigma F}\right)
_{\rho\theta}{}^{\chi\nu}F_{\mu\xi}\square F_{\chi\nu}\label{2daordem}\\
&  \hspace{-1cm}-\frac{\lambda^{2}}{24\pi^{2}\epsilon}\int d^{4}x~\left(
K_{\sigma F}\right)  ^{\mu\xi\eta\theta}\left(  K_{\sigma F}\right)  _{\theta
}{}^{\lambda\chi\nu}F_{\mu\xi}\left(  \partial_{\eta}\partial_{\lambda
}\right)  F_{\chi\nu}.\nonumber
\end{align}
We can note that the first term is a dimension-four operator while the two
last are dimension-six operators.

Finally, we can show starting from Eqs. (\ref{pol1},\ref{pol2},\ref{pol3})
that the divergent contributions to the vacuum polarization tensor are purely
transversal.\ Thus, a direct verification yields,
\begin{equation}
q_{\nu}\Pi^{\mu\nu}\left(  q\right)  =0, \label{qPi}%
\end{equation}
It assures, at 1-loop level, the absence of gauge anomalies\ and,
consequently,\ gauge symmetry preservation in the context of the modified QED
of Lagrangian (\ref{L1a}).

\section{Conclusions and Perspectives}

In this work, we have studied the contributions to the effective action of the
electromagnetic field induced by the dimension-five nonminimal coupling
$\lambda\left(  K_{\sigma F}\right)  _{\mu\nu\alpha\beta}F^{\alpha\beta}%
\bar{\psi}\sigma^{\mu\nu}\psi$ with the tensor $\left(  K_{\sigma F}\right)
_{\mu\nu\alpha\beta}$ having the same symmetries of the $(K_{F})_{\mu\nu
\rho\sigma}$. Specifically, we have focused in the quadratic gauge field terms
generated by the 1-loop radiative corrections. Our main result is that such
contributions have generated the CPT-even term of the SME electrodynamics
$(K_{F})_{\mu\nu\rho\sigma}F^{\mu\nu}F^{\rho\sigma}$. Furthermore, at second
order in $\left(  K_{F}\right)  $, other CPT-even terms containing
fourth-order derivatives involving dimension-six operators were also generated.

Our analysis has shown that the first-order LV correction to the one-loop
vacuum polarization leads to the dimension-four CPT-even term of the SME,
$\left(  me\lambda/8\pi^{2}\right)  (K_{\sigma F})_{\mu\nu\rho\sigma}F^{\mu
\nu}F^{\rho\sigma},$\ as presented in Eq. (\ref{EffA2}). This implies that the
Maxwell electrodynamics must be modified by the inclusion of such term in its
structure,\ implying a new LV electrodynamics ruled by the Lagrangian
density\ $-\frac{1}{4}F^{\mu\nu}F_{\mu\nu}-\frac{1}{4}g\lambda(K_{\sigma
F})_{\mu\nu\rho\sigma}F^{\mu\nu}F^{\rho\sigma}$,\ where $g=me/2\pi^{2}$. As a
consequence, we can use the same phenomenology that allows to constrain the
coefficients of the tensor $\left(  K_{F}\right)  $\ with stringent upper
bounds \cite{KM1,Risse,Cherenkov2,KRussell} \ to improve the bounds on the
magnitude of the quantity $\lambda\left(  K_{\sigma F}\right)  _{\mu\nu
\alpha\beta}$\ by the factor $1/g\sim4\times10^{-4}$\ (in the electron
case)$.$\ This means that a known upper-bound for the nonbirefringent
components, $\left\vert K_{F}\right\vert <10^{-17}$, would lead to an
upper-bound as tight as $\left\vert \lambda_{e}\left(  K_{\sigma F}\right)
\right\vert <10^{-21}\left(  \text{eV}\right)  ^{-1}$\ for the\ corresponding
nonminimal coupling to the electron-photon interaction. A similar argument can
be used to transfer the existing bounds of the birefringent components,
$\left\vert K_{F}\right\vert <10^{-37},$\ to level of $\left\vert \lambda
_{e}\left(  K_{\sigma F}\right)  \right\vert <10^{-41}\left(  \text{eV}%
\right)  ^{-1}$\ while associated with the same electron-photon nonminimal interaction.

Therefore, this analysis allows to improve the previous upper bounds on
$\left\vert \lambda_{e}K_{\sigma F}\right\vert $ attained in Ref.
\cite{Frede2}, by the factors $10^{5}$ and $10^{25}$, concerning to the
nonbirefringent and birefringent sectors of the nonminimal coupling,
respectively. The higher-order derivative terms in Eq. (\ref{2daordem}) do not
lead to improvements of the upper limits already attained at first order,
since are second order terms in the $(K_{F})$ tensor. An important fact must
be noted, since the counterterm depends on the particle mass, it is reasonable
to suppose that the magnitude of the nonminimal coupling may also depend on
the particle mass under analysis, fact also remarked in Ref. \cite{Frede2}, in
which different upper bounds were stated for the electron and proton
nonminimal interactions with the electromagnetic field. In respect to the
proton nonminimal interaction, previously discussed in Ref. \cite{Frede2}, the
bounds could now be improved to the level $\left\vert \lambda_{p}\left(
K_{\sigma F}\right)  \right\vert <10^{-25}\left(  \text{eV}\right)  ^{-1}%
$\ and $\left\vert \lambda_{p}K_{\sigma F}\right\vert <10^{-44}\left(
\text{eV}\right)  ^{-1}$, for the nonbirefringent and birefringent components, respectively.

Some additional\ issues still remain under investigation. The
renormalizability of this model constitutes a sensitive question for its
physical consideration as a sounder theoretical alternative. For it, the
1-loop electron self-energy and vertex corrections jointly with the 1-loop
vacuum polarization should be evaluated and analyzed. The electron self-energy
evaluation provides the usual QED contribution and others coming from the
nonminimal coupling, as the following term to the fermion effective action:
\begin{equation}
\ \bar{\psi}i\kappa^{\mu\nu}\gamma_{\mu}\partial_{\nu}\psi d^{4}%
x,\label{1.26)}%
\end{equation}
where $\kappa_{\mu\nu}=(K_{\sigma F})^{\beta}{}_{\mu\beta\nu}$\ characterizes
the nonbirefringent part of $(K_{\sigma F})_{\mu\nu\rho\sigma}$. This tensor
may be identified with\ the SME tensor $c^{\mu\nu},$ which is associated with
the fermion contribution, $\bar{\psi}c^{\mu\nu}\gamma_{\mu}\partial_{\nu}\psi
$, once only the symmetric part \cite{Colladay2} of the tensor $c^{\mu\nu}%
$\ is physically relevant. The vertex corrections involve different or new
aspects not considered in the other vertex corrections known in the
Lorentz-violating literature\ \cite{Kostelecky2002,Carone,Chen}. So, at first
order in $(K_{\sigma F})_{\mu\nu\alpha\beta}$ tensor, the 1-loop vertex
corrections provide new interactions, as the counterterm\ %

\begin{equation}
i\bar{\psi}\left(  \kappa^{\mu}{}_{\alpha}\sigma_{\lambda\mu}+\kappa^{\mu}%
{}_{\lambda}\sigma_{\alpha\mu}\right)  A^{\lambda}\partial^{\alpha}%
\psi,\nonumber
\end{equation}
which represents a new dimension-five\ nonminimal coupling. These evaluations
and the 1-loop renormalization analysis are under finalization.

Regardless the 1-loop renormalization of this model, it can be considered as a
low-energy effective model, once an ultraviolet cut-off is adopted. In such a
case, higher orders become irrelevant and the nonrenormalizability becomes a
nonessential issue, which justifies the proposal of such a model as a
preliminary theoretical option.

\begin{acknowledgments}
The authors are grateful to CNPq, CAPES and FAPEMA (Brazilian research
agencies) for invaluable financial support.{}
\end{acknowledgments}

\end{document}